\documentclass[12pt,a4paper]{article}
\usepackage{amssymb}
\usepackage{cite}
\usepackage{epsf}
\newlength{\HFPP}       \HFPP5.4mm
\def\preprint#1#2{\noindent\hbox{#1}\hfill\hbox{#2}\vskip 10pt}

\begin{document}
\begin{titlepage}
\def\thefootnote{\fnsymbol{footnote}}

\preprint{ITP-UH-02/98}{February 1998}
\vfill

\begin{center}
  {\Large\sc Correlation functions in the Calogero--Sutherland model
  with open boundaries}
\vfill

{\sc Holger Frahm}\footnote{e-mail: frahm@itp.uni-hannover.de}\\
{\sl
  Institut f\"ur Theoretische Physik, Universit\"at Hannover\\
  D-30167~Hannover, Germany}\\
\vspace{0.6em}
        and\\
\vspace{0.6em}
{\sc Sergey I. Matveenko}\footnote{e-mail: matveen@landau.ac.ru}\\
{\sl
  Landau Institute for Theoretical Physics\\
  Kosygina Street 2, 117940 Moscow, Russia}
\vspace{1.2em}

\end{center}
\vfill

\begin{quote}
Calogero-Sutherland models of type $BC_N$ are known to be relevant to
the physics of one-dimensional quantum impurity effects.  Here we
represent certain correlation functions of these models in terms of
generalized hypergeometric functions.  Their asymptotic behaviour
supports the predictions of (boundary) conformal field theory for the
orthogonality catastrophy and Friedel oscillations.
\end{quote}

%
\vfill
\vspace*{\fill}
\setcounter{footnote}{0}
\end{titlepage}

\section{Introduction}
One-dimensional models with inverse square interactions have attracted
considerable interest in recent years.  For the Calogero Sutherland
(CS) models
\cite{Calogero,Sutherland,olpe:83}
describing particles moving on a continuous line the many-body ground
state wave function is of Jastrow type and excitations can be written
as a product of this pair product wave function and certain
polyniomials in the coordinates.  Based on this observation the
eigenvalues of the Hamiltonian can be found by means of an asymptotic
Bethe Ansatz (ABA) solution \cite{Sutherland}.
Finite-size scaling analysis of the excitation spectrum and
predictions of conformal field theory (CFT) have been used to study
the critical behaviour of the CS model, leading to the identification
of the universality class of the model with periodic boundary
conditions as Luttinger liquid, i.e.\ a Gaussian model with central
charge $c=1$ \cite{hald:91c,kaya:91b}.  An interesting property of th
CS models is that the compact form of the eigenstates allows for an
explicit calculation of certain correlation functions
\cite{forr:92b,forr:95a,forr:95b,leps:95}, thus allowing to compare
the asymptotic predictions of CFT to exact expressions derived in a
microscopic model.

In addition to the models with periodic boundary conditions there
exists a class of CS models lacking translational invariance, in
particular the model of $BC_N$-type which is invariant under the
action of the Weyl group of type $B_N$ \cite{olpe:83}.  Again the
ground state can be written in a compact form of Jastrow type, and the
spectrum of excitations can be found \cite{yama:94,beps:95}.  An
analysis of the finite size corrections to the energies shows that the
spectrum acquires contributions due to the `boundaries' of the system
and the low-energy critical behaviour is described by a $c=1$ boundary
CFT \cite{yaky:96}.  Again, the existence of a `simple' expression for
the ground state wave function opens the possibility to compare the
predictions of boundary CFT for the asymptotic behaviour of
correlation functions with exact results.  These predictions are of
great interest at present due to the possibility to extract observable
properties of quantum impurity systems from finite size spectra (see
e.g.\ \cite{affl:94,aflu:94,affl:97}).

In the present paper we shall consider the $BC_N$-type CS model and
compute matrix elements which allow to compare the exponents
associated with the Anderson's `orthogonality catastrophy', i.e.\ the
dependence of the overlap between ground states for \emph{different}
boundary conditions on the system size, and the asymptotic behaviour
of (Friedel) density oscillations due to the existence of the boundary
with the corresponding expressions for the Luttinger liquid
\cite{aflu:94,bama:95,affl:97,befr:97,eggr:95,schm:96,lesa:97}.

First we will give a brief review of the properties of the CS model of
$BC_N$-type as well as the predictions of boundary CFT relevant to the
results of this paper.  A form of the Hamiltonian in a finite geometry
especially convenient for our studies is \cite{yaky:96}
\begin{eqnarray}
 {\cal H} &=& -\sum_{j=1}^N {\partial^2\over\partial q_j^2}
	   + 2\lambda(\lambda-1) \left({\pi\over2L}\right)^2
	     \sum_{j<k} \left\{
		  {1\over\sin^2{\pi\over2L}\left(q_j-q_k\right)}
		+ {1\over\sin^2{\pi\over2L}\left(q_j+q_k\right)}
		\right\}
\nonumber\\
	   &&+ \mu(\mu-1) \left({\pi\over2L}\right)^2
	     \sum_{j=1}^N {1\over\sin^2{\pi\over2L}q_j}
	   + \nu(\nu-1) \left({\pi\over2L}\right)^2
	     \sum_{j=1}^N {1\over\cos^2{\pi\over2L}q_j}\ .
\label{hamil}
\end{eqnarray}
Here $\lambda$, $\mu$, $\nu$ are positive coupling constants.  The
particles move on the interval $0\le q_j\le L$.  The two particle
interaction terms consist of the usual inverse square interaction of
particles moving on one half of the circle with circumfence $2L$ plus
a term from the interaction of the particle at $q_j$ with the mirror
image $-q_k\equiv 2L-q_k$ of the particle at $q_k$.  The last two
terms can be regarded as impurity potentials situated at the edges
$q=0$ ($q=L$) of the system with strength given by $\mu$ ($\nu$).
Eigenvalues and eigenstates of (\ref{hamil}) have been determined
explicitely \cite{yama:94,beps:95}, the many-particle ground state
wavefunction is \cite{olpe:83}
\begin{equation}
  \Psi_0^{(\lambda)}(q_1,\ldots,q_N;\mu,\nu) = 
	\prod_{j<k} 
	  \left|\sin{\pi\over2L}\left(q_j-q_k\right)\,
		\sin{\pi\over2L}\left(q_j+q_k\right)\right|^{\lambda}\,
	\prod_{\ell=1}^N \left|\sin{\pi\over2L}q_\ell\right|^\mu
			 \left|\cos{\pi\over2L}q_\ell\right|^\nu
\label{gs}
\end{equation}
As for the periodic CS models the spectrum can be reproduced exactly
by means of the asymptotic Bethe Ansatz method \cite{yama:94}.
Expanding the ground state energy in inverse powers of the system size
the following finite size scaling form is found \cite{yaky:96}
\begin{eqnarray}
   E_N^{(0)} &=& L \epsilon^{(0)} + 2f
             + {2\pi v_F\over L}{\lambda\over4}\left(\Delta N_b\right)^2
		- {\pi v_F\over24 L}\lambda
\label{fse0}
\\
   \Delta N_b &=& {1\over2\lambda} \left(\mu+\nu-\lambda\right)
\nonumber
\end{eqnarray}
where $\epsilon^{(0)}$ and $f$ are the bulk energy density and the
boundary energy in the thermodynamic limit for fixed particle density
$n=N/L$, respectively.  $v_F=2\pi\lambda n$ is the Fermi velocity of
the elementary excitations.  Similarly, the energy of an excited state
with $\Delta N$ additional particles and $N_{ph}>0$ particle-hole
excitations near the Fermi point is, to leading order in $1/L$,
\begin{equation}
   E - E_{N}^{(0)} - \mu_c^{(0)}\Delta N \simeq
	{2\pi v_F\over L}\left( 
	  {\lambda\over4} \left(\Delta N + \Delta N_b\right)^2
	- {\lambda\over4} \left(\Delta N_b\right)^2 \right)
	+ {\pi v_F \over L} N_{ph}\ .
\label{fsee}
\end{equation}
Here we have absorbed a term $\mu_c^{(0)}\Delta N$ into the definition
of the energy the system with $\mu_c^{(0)}=L\partial\epsilon^{(0)}/
\partial N$ being the chemical potential.  

The expressions (\ref{fse0}), (\ref{fsee}) should be compared with the
corresponding prediction of CFT for models with free boundary
conditions, namely
\begin{equation}
   E^{(x)} = L \epsilon^{(0)} + 2f - {\pi v_F\over24L} c
	+ {\pi v_F\over L} x
\end{equation}
with the Virasoro central charge $c$ appearing in the universal
amplitude of the $1/L$-term and the critical exponent $x$ of the
operator generating this state.  It is well known that the long-range
nature of the interactions in CS models gives rise to non-universal
$1/L$-contributions to the ground state energy of these systems
\cite{kaya:91b,yaky:96}.  The $\lambda$-dependence in the last term in
(\ref{fse0}) is believed to be a consequence of this effect, thus
yealding an incorrect value for the central charge.

The other contribution of order $1/L$ to the ground state energy is a
consequence of the scattering due to the free boundary and the
impurity potentials.  If one applies boundary CFT to obtain the
surface critical exponents controlling the asymptotic behaviour of
correlation functions one has to distinguish operators connecting
states corresponding to \emph{different} boundary conditions and
operators inducing a change of particle number or particle-hole
excitations in the ground state corresponding to a given boundary
condition \cite{aflu:94,affl:97}.  For the latter case the phase shift
$\Delta N_b$ should be absorbed into the the change of the number of
particles
\begin{equation}
   \tilde{E}_N^{(0)} = {E}_N^{(0)} 
	- {2\pi v_F\over L}{\lambda\over4}\left(\Delta N_b\right)^2\ ,
   \qquad
   \widehat{\Delta N} = \Delta N + \Delta N_b
\label{shift0}
\end{equation}
to restore particle hole symmetry of the finite size spectrum
(\ref{fsee}).  The resulting scaling dimension of an operator $\phi$
corresponding to this situation is
\begin{equation}
   x(\phi) = {L\over\pi v_F} \left( E-\tilde{E}_N^{(0)} \right)^2
	  = {\lambda\over2} \left( \widehat{\Delta N} \right)^2 + N_{ph}
\label{exp:bb}
\end{equation}
with integer $\widehat{\Delta N}$.

To obtain the conformal dimension of boundary condition changing
operators finite size energies corresponding to states subject to
\emph{different} boundary conditions have to be compared
\cite{aflu:94,affl:97,befr:97}.  Consequently, only one of the two
phase shifts $\Delta N_b$, $\Delta N_{b'}$ arising in these
expressions can be absorbed into a shift of the particle number as in
(\ref{shift0}).  For the operator $\psi_{bb'}$ connecting the gound
states corresponding to different boundary conditions this leads to an
operator dimension
\begin{equation}
   x(\psi_{bb'}) = {\lambda\over2} 
	\left( \Delta N_b - \Delta N_{b'} \right)^2\ .
\label{exp:bc}
\end{equation}
This exponent determines the orthogonality exponent
$\langle0_b|0_{b'}\rangle \propto L^{-x}$ and the related X-ray edge
singularity arising from a sudden change of the boundary potential
\cite{aflu:94,affl:97,befr:97,esfr:97}.

The asymptotic behaviour of correlation functions in the bulk is still
determined by the conformal dimension of the corresponding operator.
In fact, an $n$-point function of the semiinfinite system is subject
to different boundary conditions but obeys the same differential
equation as the $2n$-point function including the mirror positions in
the system without boundary \cite{card:84}.  Hence, the critical
exponent for the asymptotic behaviour of the single particle density
$\langle\rho(q)\rangle$ due to the presence of the boundary can be
obtained from the density density correlation function for the
corresponding system without a boundary, namely the periodic CS model.
The most dominant term is due to backscattering processes with
momentum $\pm 2k_F\equiv 2\pi n$ and decays asymptotically as
\cite{kaya:91b}:
\begin{equation}
   \langle\rho(q)\rangle - n \sim
	{\cos(2 k_F q) \over q^{1/\lambda}}.
\label{friedel}
\end{equation}

\section{Overlap integrals}
To compute the overlap integral between ground states of (\ref{hamil})
corresponding to different values of the boundary field stengths
$\mu$, $\nu$ we will make use their representation in terms of so
called Selberg correlation integrals (see \cite{forr:92b} and
references therein)
\begin{equation}
  S_{n,m}(\lambda_1,\lambda_2,\lambda; x_1,\ldots x_m) =
   \left(\prod_{\ell=1}^n 
	 \int_0^1 dt_\ell\ \prod_{p=1}^m (t_\ell-x_p) \right)
	 D_{\lambda_1,\lambda_2,\lambda}(t_1,\ldots,t_n)
\label{selbi}
\end{equation}
with
\begin{equation}
   D_{\lambda_1,\lambda_2,\lambda}(t_1,\ldots,t_n) =
    \prod_{\ell=1}^n t_\ell^{\lambda_1-1} (1-t_\ell)^{\lambda_2-1}
    \prod_{1\le j<k\le n} \left| t_k-t_j \right|^\lambda\ .
\end{equation}
For $m=0$ the integrals can be evaluated with result expressed in
terms of Gamma-functions
\begin{equation}
   S_{n,0}(\lambda_1,\lambda_2,\lambda) =
    \prod_{j=1}^n 
    {{\Gamma(1+\lambda j/2)\Gamma(\lambda_1+\lambda(j-1)/2)
	\Gamma(\lambda_2+\lambda(j-1)/2)} \over
     {\Gamma(1+\lambda/2)\Gamma(\lambda_1+\lambda_2+\lambda(n+j-2)/2)}
    }\ .
\label{selbn0}
\end{equation}
We first note that the normalization integral of (\ref{gs})
\begin{equation}
   {\cal N}_{\lambda,\mu,\nu} =
     \left( \prod_{\ell=1}^N \int_0^L dq_\ell \right)
	|\Psi_0^{(\lambda)}(q_1,\ldots,q_N;\mu,\nu)|^2
\end{equation}
is of the form (\ref{selbi}):  Substituting $t_\ell =
{1\over2}\left(1-cos{\pi\over L}q_\ell \right)$ we obtain
\begin{eqnarray}
   {\cal N}_{\lambda,\mu,\nu} &=&
	\left(\prod_{\ell=1}^N {L\over2\pi} \int_0^1 dt_\ell\,
		t_\ell^{\mu-{1\over2}} (1-t_\ell)^{\nu-{1\over2}}
	\right)
	\prod_{1\le j<k\le n} \left| t_k-t_j \right|^{2\lambda}
\nonumber\\
   &=& \left({L\over2\pi}\right)^N 
	S_{N,0}\left(\mu+{1\over2},\nu+{1\over2},2\lambda\right)\ .
\end{eqnarray}
Now the overlap between two states (\ref{gs}) for  different sets of
boundary fields $(\mu,\nu)$ and $(\mu',\nu')$ can be expressed as the
ratio of Selberg correlation integrals
\begin{equation}
   \left| \langle \mu,\nu| \mu',\nu'\rangle \right|^2 =
    {\left(S_{N,0}((\mu+\mu'+1)/2,(\nu+\nu'+1)/2,2\lambda)\right)^2
     \over
      S_{N,0}(\mu+1/2,\nu+1/2,2\lambda)
      S_{N,0}(\mu'+1/2,\nu'+1/2,2\lambda)}\ .
\end{equation}
Using (\ref{selbn0}) we obtain
\begin{eqnarray}
   \ln|\langle \mu,\nu| \mu',\nu'\rangle| =
      {1\over2} \sum_{j=0}^{N-1} \Big(
	f(\mu,\mu',{1\over2}+\lambda j) +
	f(\nu,\nu',{1\over2}+\lambda j) \qquad
\nonumber\\
	- f(\mu+\nu,\mu'+\nu',1 +\lambda (N+j-1)) \Big)
\label{logovl}
\end{eqnarray}
with
\begin{equation}
   f(\mu,\mu',x) =
   \ln {\Gamma^2\left({1\over2}(\mu+\mu') + x\right) \over
    \Gamma\left(\mu + x\right)\Gamma\left(\mu' + x\right)}
    \approx \left(\mu-\mu'\right)^2 
	\left(-{1\over4x} +{1\over8x^2}\left(\mu+\mu'-1\right)
		+\cdots\right)\ .
\end{equation}
For large system size $L$ and fixed density $\rho=N/L$ the overlap
integral (\ref{logovl}) is determined by the logarithmic divergence of
the first two contributions to the sum at $j=N-1$ giving
\begin{equation}
  \ln \left|\langle \mu,\nu| \mu',\nu'\rangle\right|
	\propto {1\over8\lambda}
	\left((\mu-\mu')^2 + (\nu-\nu')^2\right)\, \ln L
\end{equation}
which is in perfect agreement with (\ref{fse0}) and the prediction
(\ref{exp:bc}) of boundary CFT (note that the operator connecting
$|\mu,\nu\rangle$ and $|\mu',\nu'\rangle$ in this situation is a
product of two boundary changing operators $\psi_{\mu\mu'}$ and
$\psi_{\nu\nu'}$ acting at the boundary at $q=0$ and $q=L$,
respectively \cite{befr:97}).

\section{Friedel oscillations}
\begin{figure}
\caption{Single particle density oscillations for a system of $N=10$
  particles with $\lambda=2$, $\mu=3$ and $\nu=1$.
\label{fried10}}
\begin{center}
\epsfxsize=\textwidth
\epsffile{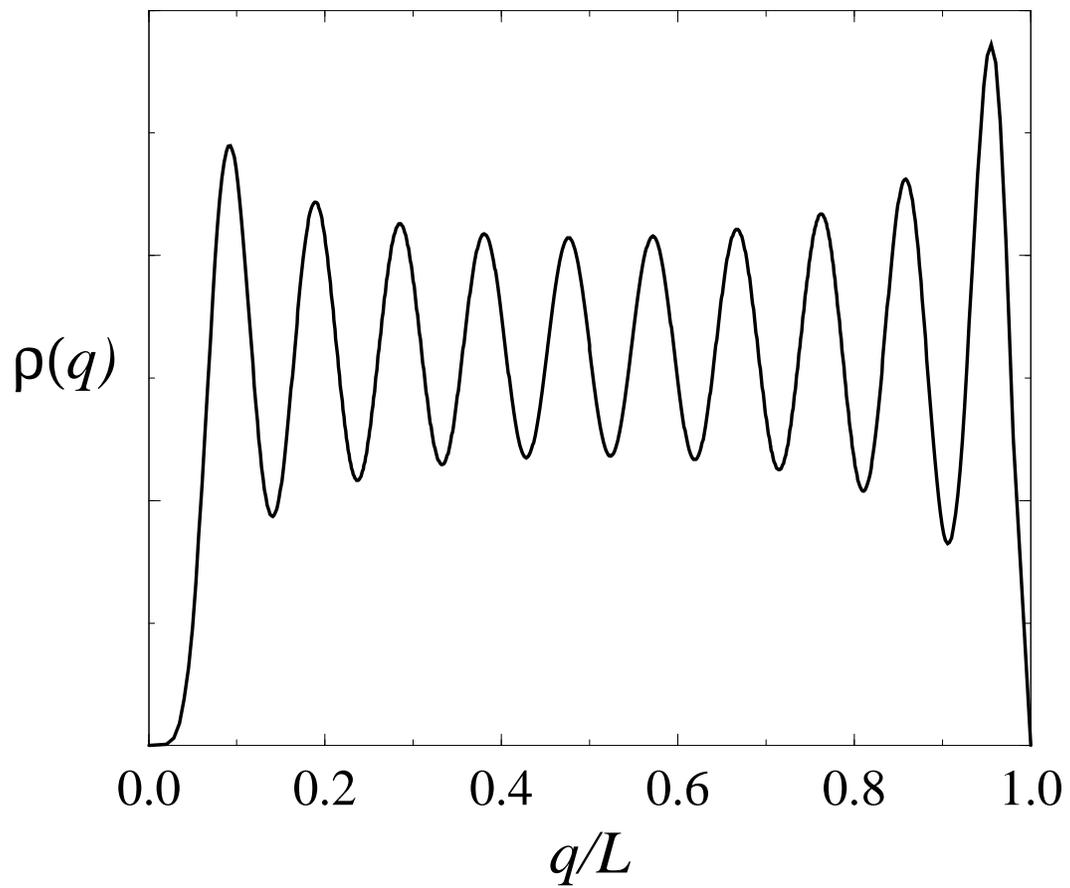}
\end{center}
\end{figure}

We consider now a spatial dependence for single particle density
\begin{equation}
   \langle\rho(q)\rangle = \frac{1}{{\cal N}_{\lambda,\mu,\nu}}  
	\left( \prod_{\ell=2}^N\int_0^L dq_\ell \right)
	|\Psi_0^{(\lambda)}(q, q_2, \ldots,q_N;\mu,\nu)|^2.
\label{r1}
\end{equation}
For integer values $\lambda$ the function $\langle\rho(q)\rangle$ can
be expressed in terms of the Selberg integrals (\ref{selbi})
\begin{equation}
   \langle\rho(q)\rangle = 
	\frac{2\pi}{L}x^{\mu} (1-x)^{\nu}\frac{S_{N-1,2\lambda}
	\left(\mu+\frac{1}{2},\nu+\frac{1}{2},2\lambda; x_1=\ldots =x_m\right)}
	{S_{N,0}\left(\mu+\frac{1}{2},\nu+\frac{1}{2},2\lambda\right)}\ ,
\label{r2}
\end{equation}
where $m=2\lambda$ and $x_1=\ldots=x_m=x=\sin^2({\pi q}/{2L})$.
Eq.~(\ref{r2}) can be rewritten as
\begin{eqnarray}
   \langle\rho(q)\rangle 
	&=& \frac{2\pi}{L}x^{\mu}(1-x)^{\nu}
	\frac{S_{N-1,0}(\mu+\frac{1}{2}+2\lambda,
		\nu+\frac{1}{2},2\lambda)}
	     {S_{N,0}(\mu+\frac{1}{2},\nu+\frac{1}{2},2\lambda)} 
\nonumber \\
   &&\times\,
	{^{}_2}F_1^{(\lambda)}(-N+1,
		\frac{1}{\lambda}(\mu+\nu+m)+N-2, 
		\frac{1}{\lambda}(\mu-\frac{1}{2}+m),
		x_1=\ldots =x_m ),
\label{r3}
\end{eqnarray}
where ${^{}_2}F_1^{(\lambda )}$ is a generalized hypergeometric
function of $m$ variables (see \cite{forr:92b}).  For finite systems
this expression for the single particle density can be evaluated by
using the fact that ${^{}_2}F_1^{(\lambda )}$ for equal arguments can
be written in tems of Jack symmetric polynomials
\cite{macdonald:sympoly}.  In Fig.~\ref{fried10} we have plotted
(\ref{r3}) in this representation for a system of $N=10$ particles.

In the thermodynamic limit we derive the asymptotic behavior of
$\langle\rho(q)\rangle$ for $1\ll q\ll N$ using the integral
representation \cite{forr:95a}
\begin{eqnarray}
  &&{^{}_2}F_1^{(2/\lambda )} (a, \lambda_1 +\lambda (m-1)/2, \lambda_1 +
  \lambda_2 +\lambda (m-1), x_1=\ldots =x_m )
\nonumber \\
   &&\qquad
   =\left(S_{m,0} (\lambda_1,\lambda_2,\lambda)\right)^{-1}
      \left(\prod_{\ell=1}^m \int_{-\infty}^0 dt_\ell(1-xt_l )^{-a}\right)
      D_{\lambda_1,\lambda_2, \lambda } (t_1,\ldots ,t_m )\ .
\label{r4}
\end{eqnarray}
Omitting the $x$-independent factor we find from (\ref{r3})
\begin{equation}
   \langle\rho(q)\rangle 
   \propto x^{\mu }(1-x)^{\nu }
	\prod_{\ell=1}^m
	 \left(\int_0^{\infty} dt_\ell 
		\left\{\frac{(1+xt_\ell)t_\ell}{(1+t_\ell)}
		\right\}^{\bar{n}}
	\frac{t_\ell^{(\mu +\nu +1)/\lambda -2}}
	{(1+t_\ell)^{2+(\nu -\frac{1}{2})/\lambda}}\right)
	\prod_{j<k}^m \vert t_j -t_k\vert^{2/\lambda}\ ,
\label{r5}
\end{equation}
where $\bar{n}=N-1$.  Near the boundary $q\ll 1$ we obtain
\begin{equation}
   \langle\rho(q)\rangle \propto q^{2\mu }\ .
\end{equation}

For $1\ll q\ll N$ we have $x =\sin^2(\pi q/2L)\to g^2/\bar{n}^2$,
$g=k_F q/2$.  After rescaling $t_l \to \bar{n}t_l$ and using the
identity $\lim_{n\to\infty}{\left(1+y/n\right)^n}=\exp y$ we obtain
from (\ref{r5})
\begin{equation}
   \langle\rho(q)\rangle \propto q^{2\mu}
	\left(\prod_{\ell=1}^m\int_0^{\infty}dt_\ell 
	\exp{(g^2 t_\ell -1/t_\ell)} 
	t_\ell^{\frac{3}{2\lambda} -4 +\frac{\mu}{\lambda}}\right)
	\prod_{j<k} \vert t_j - t_k \vert^{2/\lambda}.
\end{equation}
In order to calculate this divergent integral we analytically continue
it to imaginary $g$ and obtain after rescaling $t \to t/g $
\begin{equation}
   \langle\rho(q)\rangle \propto g^{m -1}
	\left(\prod_{\ell=1}^m \int_0^{\infty}dt_\ell
	\exp^{-g(t_\ell+\frac{1}{t_\ell})} 
	t_\ell^{\frac{3}{2\lambda}-4+\frac{\mu}{\lambda}}\right)
	\prod_{j<k}\vert t_j - t_k\vert^{{2}/{\lambda}}.
\label{r7}
\end{equation}
Now Eq.~(\ref{r7}) can be easy evaluated in the limit $q \gg L/N$ which
corresponds to $g\gg1$.  Considering the case $m=2$ first we find
after an integration near the extremum point $t_j = 1$
\begin{equation}
   \langle\rho(q)\rangle
   \propto \frac{\cos 2 k_F q}{q^{1/{\lambda}}}.
\label{r}
\end{equation}
In the general case the main oscillating term is obtained by
integration (\ref{r7}) over an region where only two variables are
near extremum points.  Then the integration over these two variables
gives rise to a factor
\[
   \frac{\cos 2k_F q}{q^{1/{\lambda }+1}}
\]
while the other integrals of type $\prod \int e^{-gt}f(t)$ contribute
a factor of order $(\frac{1}{g})^{m-2}$.  Substituting to Eq.\
(\ref{r7}) we reproduce the result (\ref{r}) for any value of the
coupling constant $\lambda$.  Again, this result is in complete
agreement with the boundary CFT prediction (\ref{friedel}).

\section{Free fermionic case}
Note that in the free fermionic case $\lambda =1$ the Friedel
oscillations can be investigated in more detail due to a possibility
to express the Selberg integral $S_{n,2}$ in terms of Appell's
hypergeometric function $F_4$ \cite{forr:92b}
\begin{eqnarray}
   &&S_{n,2} (\lambda_1,\lambda_2 ,\lambda ;x_1, x_2 )
\nonumber \\
   &&\qquad =
	(-1)^n S_{n,0}(\lambda_1+1,\lambda_2+1,\lambda)
	F_4(a,b,c-1,c_2;(1-x_1)(1-x_2)),
\label{F4}
\end{eqnarray}
where
\begin{equation}
  a=-{n},\, 
  b=\frac{2}{\lambda } (\lambda_1 + \lambda_2 +1)+{n}-1,\, 
  c_1 =\frac{2\lambda_1 }{\lambda },\, 
  c_2 =\frac{2\lambda_2 }{\lambda }.
\end{equation}
Substituting (\ref{F4}) to eq.(\ref{r2}) we find
\begin{equation}
  \langle\rho (q)\rangle \propto x^{\mu }(1-x)^{\nu } 
	F_4 (-\bar{n},\mu+\nu+1+{\bar{n}},\mu +\frac{1}{2}, 
		\nu +\frac{1}{2}, x^2, (1-x)^2 ),
\label{r11}
\end{equation}
where $x = \sin^2({k_F q}/{2N})$ as before.  The Appell function
$F_4 (\alpha , \beta , \gamma_1 , \gamma_2 ,x,y)$ satisfies the
following system of equations
\begin{eqnarray}
  &&x(1-x) Z^{\prime \prime }_{xx} -y^2 Z^{\prime \prime }_{yy}
  -2xyZ^{\prime \prime }_{xy}  +[\gamma_1 -(\alpha +\beta +1 )x ]
  Z^{\prime }_{x} -(\alpha +\beta +1 )y Z^{\prime }_{y} 
  -\alpha \beta Z =0,
\nonumber\\
\label{zz}\\
  &&y(1-y) Z^{\prime \prime }_{yy} -x^2 Z^{\prime \prime }_{xx}
  -2xyZ^{\prime \prime }_{xy}+[\gamma_2 -(\alpha +\beta +1 )y ]
  Z^{\prime }_{y} -(\alpha +\beta +1 )x Z^{\prime }_{x} -\alpha \beta Z =0.
\nonumber
\end{eqnarray}
In the limit $q \ll N $ we obtain after simple manipulations
\begin{equation}
  \langle\rho (q)\rangle \propto Z (u,v) \vert_{u = 2q^2, v=0 },
\end{equation}
where  $Z(u,v)$ is the solution of equations
\begin{eqnarray}
  &&Z^{\prime \prime }_{uv} = \frac{\mu (\mu -1 )Z}{(u+v)^2 },
\nonumber\\
\label{u}\\
  &&2uZ^{\prime \prime }_{uu} -2vZ^{\prime \prime }_{vv}
  -2(u-v)Z^{\prime \prime }_{uv}+3Z^{\prime }_v +k_F^2 Z = 0.
\nonumber
\end{eqnarray}
For $u \gg 1$ in the first order variables $u$ and $v$ are separated
each from other. We have as a result
\begin{equation}
  \langle\rho (q)\rangle
	\propto n + \frac{1}{q^{1/2}}Z_{\delta }(2 k_F q ),
\label{be}
\end{equation}
where $Z_{\delta }$ is the Bessel function and
\[
   \delta = \sqrt{1/4 + 8\mu (\mu -1)}.
\]
The asymptotic behavior of (\ref{be}) is in agreement with general
result (\ref{r}).

\section{Summary and conclusion}
We have studied the asymptotic behaviour of the overlap integral
between ground states of the $BC_N$ type Calogero Sutherland model
corresponding to different strenghts of the boundary fields and the
Friedel oscillations of the single particle density due to the
presence of a boundary.  Our results are in agreement with those
obtained by applying the predictions of boundary conformal field
theory to the finite size spectra of these systems.  In particular, it
has been established that non universal exponents showing a continuous
dependence on the values of the boundary fields can arise in the
correlation functions of boundary changing operators.

\section*{Acknowledgements}
This work has been supported by the Deutsche Forschungsgemeinschaft under
Grant No.\ Fr~737/2--2.  S.M.\ thanks Institut f\"ur Theoretische Physik,
Universit\"at Hannover for kind hospitality and the Russia  Foundation of
Fundamental Research for support under Grant No.~960217791.

\newpage
\setlength{\baselineskip}{14pt}


\begin{thebibliography}{10}

\bibitem{Calogero}
F. Calogero, J. Math. Phys. {\bf 10},  2191 and 2197  (1969);
{\em ibid.} {\bf 12},  419  (1971).

\bibitem{Sutherland}
B. Sutherland, J. Math. Phys. {\bf 12},  246 and 251 (1971);
Phys. Rev. A {\bf 4},  2019  (1971);
{\em ibid.} {\bf 5},  1372  (1972).

\bibitem{olpe:83}
M.~A. Olshanetsky and A.~M. Perelomov, Phys. Rep. {\bf 94},  313  (1983).

\bibitem{hald:91c}
F.~D.~M. Haldane, Phys. Rev. Lett. {\bf 66},  1529  (1991).

\bibitem{kaya:91b}
N. Kawakami and S.-K. Yang, Phys. Rev. Lett. {\bf 67},  2493  (1991).

\bibitem{forr:92b}
P.~J. Forrester, Nucl. Phys. B {\bf 388},  671  (1992).

\bibitem{forr:95a}
P.~J. Forrester, J. Math. Phys. {\bf 36},  86  (1995).

\bibitem{forr:95b}
P.~J. Forrester, Mod. Phys. Lett. B {\bf 9},  359  (1995).

\bibitem{leps:95}
F. Lesage, V. Pasquier, and D. Serban, Nucl. Phys. B {\bf 435 [FS]},  585
  (1995).

\bibitem{yama:94}
T. Yamamoto, J. Phys. Soc. Japan {\bf 63},  1223  (1994).

\bibitem{beps:95}
D. Bernard, V. Pasquier, and D. Serban, Europhys. Lett. {\bf 30},  301  (1995).

\bibitem{yaky:96}
T. Yamamoto, N. Kawakami, and S.-K. Yang, J. Phys. A {\bf 29},  317  (1996).

\bibitem{affl:94}
I. Affleck,  in {\em Correlation effects in low-dimensional electron systems},
  Vol.~118 of {\em Springer Series in Solid-State Sciences}, edited by A. Okiji
  and N. Kawakami (Springer Verlag, Berlin, 1994), pp.\ 82--95.

\bibitem{aflu:94}
I. Affleck and A.~W.~W. Ludwig, J. Phys. A {\bf 27},  5375  (1994).

\bibitem{affl:97}
I. Affleck, Nucl. Phys. B (Proc. Suppl.) {\bf 58},  35  (1997).

\bibitem{bama:95}
A.~V. Balatsky and S.~I. Matveenko, Phys. Rev. B {\bf 52},  R8676  (1995).

\bibitem{befr:97}
G. Bed{\"u}rftig and H. Frahm, J. Phys. A {\bf 30},  4139  (1997).

\bibitem{eggr:95}
R. Egger and H. Grabert, Phys. Rev. Lett. {\bf 75},  3505  (1995).

\bibitem{schm:96}
P. Schmitteckert and U. Eckern, Phys. Rev. B {\bf 53},  15397  (1996).

\bibitem{lesa:97}
F. Lesage and H. Saleur, J. Phys. A {\bf 30},  L457  (1997).

\bibitem{esfr:97}
F.~H.~L. E{\ss}ler and H. Frahm, Phys. Rev. B {\bf 56},  6631  (1997).

\bibitem{card:84}
J.~L. Cardy, Nucl. Phys. B {\bf 240 [FS12]},  514  (1984).

\bibitem{macdonald:sympoly}
I.~G. Macdonald, {\em Symmetric Functions and {Hall} Polynomials}, 2nd ed.
  (Clarendon Press, Oxford, 1995).

\end{thebibliography}
\end{document}